\documentclass[letterpaper]{article}
\usepackage{chicagor}
\usepackage{aaai1}
\usepackage{times}
\usepackage{helvet}
\usepackage{courier}
\usepackage{graphicx}
\usepackage{epsfig}

\newtheorem{THEOREM}{Theorem}[section]
\newenvironment{theorem}{\begin{THEOREM} \hspace{-.85em} {\bf :} }%
                        {\end{THEOREM}}
\newtheorem{LEMMA}[THEOREM]{Lemma}
\newenvironment{lemma}{\begin{LEMMA} \hspace{-.85em} {\bf :} }%
                      {\end{LEMMA}}
\newtheorem{COROLLARY}[THEOREM]{Corollary}
\newenvironment{corollary}{\begin{COROLLARY} \hspace{-.85em} {\bf :} }%
                          {\end{COROLLARY}}
\newtheorem{PROPOSITION}[THEOREM]{Proposition}
\newenvironment{proposition}{\begin{PROPOSITION} \hspace{-.85em} {\bf :} }%
                            {\end{PROPOSITION}}
\newtheorem{DEFINITION}[THEOREM]{Definition}
\newenvironment{definition}{\begin{DEFINITION} \hspace{-.85em} {\bf :} \rm}%
                            {\end{DEFINITION}}
\newtheorem{CLAIM}[THEOREM]{Claim}
\newenvironment{claim}{\begin{CLAIM} \hspace{-.85em} {\bf :} \rm}%
                            {\end{CLAIM}}
\newtheorem{EXAMPLE}[THEOREM]{Example}
\newenvironment{example}{\begin{EXAMPLE} \hspace{-.85em} {\bf :} \rm}%
                            {\end{EXAMPLE}}
\newtheorem{REMARK}[THEOREM]{Remark}
\newenvironment{remark}{\begin{REMARK} \hspace{-.85em} {\bf :} \rm}%
                            {\end{REMARK}}

\newcommand{\thm}{\begin{theorem}}
\newcommand{\lem}{\begin{lemma}}
\newcommand{\pro}{\begin{proposition}}
\newcommand{\dfn}{\begin{definition}}
\newcommand{\rem}{\begin{remark}}
\newcommand{\xam}{\begin{example}}
\newcommand{\cor}{\begin{corollary}}

\newcommand{\ethm}{\end{theorem}}
\newcommand{\elem}{\end{lemma}}
\newcommand{\epro}{\end{proposition}}
\newcommand{\edfn}{\bbox\end{definition}}
\newcommand{\erem}{\bbox\end{remark}}
\newcommand{\exam}{\bbox\end{example}}
\newcommand{\ecor}{\end{corollary}}

\newcommand{\beqn}{\begin{equation}}
\newcommand{\eeqn}{\end{equation}}

\newcommand{\bbox}{\vrule height7pt width4pt depth1pt}

\newcommand{\clm}{\begin{claim}}
\newcommand{\eclm}{\end{claim}}

\newcommand{\bor}{\bigvee}

\renewcommand{\phi}{\varphi}

\newcommand{\F}{{\cal F}}
\newcommand{\G}{{\cal G}}

\newcommand{\<}{\langle}
\renewcommand{\>}{\rangle}

\newcommand{\ol}{\setlength{\itemsep}{0pt}\begin{enumerate}}
\newcommand{\eol}{\end{enumerate}\setlength{\itemsep}{-\parsep}}
\newcommand{\ul}{\setlength{\itemsep}{0pt}\begin{itemize}}
\newcommand{\dl}{\setlength{\itemsep}{0pt}\begin{description}}
\newcommand{\edl}{\end{description}\setlength{\itemsep}{-\parsep}}
\newcommand{\eul}{\end{itemize}\setlength{\itemsep}{-\parsep}}

\newcommand{\commentout}[1]{}

\newcommand{\bi}{\begin{itemize}}
\newcommand{\ei}{\end{itemize}}
\newcommand{\be}{\begin{enumerate}}
\newcommand{\ee}{\end{enumerate}}

\newenvironment{oldthm}[1]{\par\noindent{\bf Theorem #1:} \em \noindent}{\par}
\newenvironment{oldlem}[1]{\par\noindent{\bf Lemma #1:} \em \noindent}{\par}
\newenvironment{oldcor}[1]{\par\noindent{\bf Corollary #1:} \em \noindent}{\par}
\newenvironment{oldpro}[1]{\par\noindent{\bf Proposition #1:} \em \noindent}{\par}
\newcommand{\othm}[1]{\begin{oldthm}{\ref{#1}}}
\newcommand{\eothm}{\end{oldthm} \medskip}
\newcommand{\olem}[1]{\begin{oldlem}{\ref{#1}}}
\newcommand{\eolem}{\end{oldlem} \medskip}
\newcommand{\ocor}[1]{\begin{oldcor}{\ref{#1}}}
\newcommand{\eocor}{\end{oldcor} \medskip}
\newcommand{\opro}[1]{\begin{oldpro}{\ref{#1}}}
\newcommand{\eopro}{\end{oldpro} \medskip}

\newcommand{\bxor}[1]{\dot{\bor}}

\nocopyright
\renewcommand{\chicagoraddresspub}{\commentout}

\begin{document}

\title{Beyond Nash Equilibrium:\\
Solution Concepts for the 21st Century%
\thanks{Reprinted from the \emph{Proceedings of the Twenty-Seventh
Annual ACM Symposium on Principles of Distributed Computing}, 2008.
Work supported in part by NSF under
under grants ITR-0325453 and IIS-0534064, and by AFOSR under grant 
FA9550-05-1-0055.}}
\author{Joseph Y. Halpern\\
Cornell University\\
Dept. of Computer Science\\
Ithaca, NY 14853\\
halpern@cs.cornell.edu\\
http://www.cs.cornell.edu/home/halpern}

\maketitle

\setcounter{page}{0}
\thispagestyle{empty}

\begin{abstract}
Nash equilibrium is the most commonly-used notion of equilibrium in game 
theory.  However, it suffers from numerous problems.  Some are well known 
in the game theory community; for example, the Nash equilibrium of repeated
prisoner's dilemma is neither normatively nor descriptively reasonable.
However, new problems arise when considering Nash equilibrium from a 
computer science perspective: for example, Nash equilibrium is not robust
(it does not tolerate ``faulty'' or ``unexpected'' behavior), it does not
deal with coalitions, it does not take computation cost into account, and
it does not deal with cases where players are not aware of all aspects of
the game.  Solution concepts that try to address these shortcomings of Nash
equilibrium are discussed.  
\end{abstract}

\section{Introduction}
\emph{Nash equilibrium} is the most commonly-used notion of equilibrium
in game theory.  Intuitively, a Nash equilibrium is a \emph{strategy
profile} (a collection of strategies, one for each player in the game)
such that no player can do better by deviating.  The intuition behind
Nash equilibrium is that it represent a possible steady state of play.
It is a fixed point where each player holds correct beliefs about what
other players are doing, and plays a best response to those beliefs.
Part of what makes Nash equilibrium so attractive is that in games 
where each player has only finitely many possible deterministic
strategies, and we allow mixed (i.e., randomized) strategies, there is
guaranteed to be a Nash equilibrium \cite{Nash50} (this was, in fact,
the key result of Nash's thesis).

For quite a few games, thinking in terms of Nash equilibrium gives
insight into what people do  (there is a reason that game theory is
taught in business schools!).  However, as is well known, Nash
equilibrium suffers from numerous problems.  For example, the Nash
equilibrium in games such as repeated 
prisoner's dilemma is to always defect (see
Section~\ref{sec:computation} for more 
discussion of repeated prisoner's dilemma).  It is hard to make a case
that rational players ``should'' play the Nash equilibrium in this game
when ``irrational'' players who cooperate for a while do much better! 
Moreover, in a game that is only played once, why should a Nash
equilibrium arise when there are multiple Nash equilibria?  Players
have no way of knowing which one will be played.  And even in games
where there is a unique Nash equilibrium (like repeated prisoner's
dilemma), how do players obtain correct beliefs about what other players 
are doing if the game is played only once?  (See \cite{Kreps90} for a
discussion of some of these problems.)  

Not surprisingly, there has been a great deal of work in the economics
community on developing alternative solution concepts.  Various
alternatives to and refinements of Nash equilibrium have been
introduced, including, among many others, \emph{rationalizability},
\emph{sequential equilibrium}, \emph{(trembling hand) perfect
equilibrium}, \emph{proper equilibrium}, and \emph{iterated deletion of
weakly dominated strategies}.  (These notions are discussed in
standard game theory text, such as \cite{ft:gt,OR94}.)  
Despite some successes, none of these alternative solution concepts
address the following three problems with Nash equilibrium, all inspired
by computer science concerns.
\begin{itemize}
\item Although both computer science and distributed computing are
concerned with multiple agents interacting, the focus in the game theory
literature has been on the strategic concerns of agents---rational
players choosing strategies that are best responses to strategies chosen
by other player, the focus in distributed computing has been on problems
such as fault tolerance and asynchrony, leading to, for example 
work on Byzantine agreement \cite{FLP,PSL}.  Nash equilibrium does not
deal with ``faulty'' or ``unexpected'' behavior, nor does it deal with
colluding agents.  In large games, we should expect both.
\item Nash equilibrium does not take computational concerns into
account.  We need solution concepts that can deal with resource-bounded
players, concerns that are at the heart of cryptography. 
\item Nash equilibrium presumes that players have common knowledge of the
structure of the game, including all the possible moves that can be made
in every situation and all the players in game.  This is not always
reasonable in, for example, the large auctions played over the internet.
\end{itemize}
In the following sections, I discuss each of these issues in more
detail, and sketch solution concepts that can deal with them, with
pointers to the relevant literature.

\section{Robust and Resilient Equilibrium}\label{sec:robust}

Nash equilibrium tolerates deviations by one player.  It is perfectly
consistent with Nash equilibrium that two players could do much better
by deviating in a coordinated way.  For example, consider a game with $n
> 1$ players where players much play either 0 or 1.  If everyone plays
0, everyone get a payoff of 1; if exactly two players plays 1 and the
rest play 0, then the two who play 1 get a payoff of 2, and the rest get
0; otherwise, everyone gets 0.  Clearly everyone playing 0 is a Nash
equilibrium, but any pair of players can do better by deviating and
playing 1.

Say that a Nash equilibrium is \emph{$k$-resilient} if it tolerates
deviations by coalitions of up to $k$ players.  The notion of resilience
is an old one in the game theory literature, going  back to Aumann
\citeyear{Aumann59}.  Various extensions of Nash equilibrium have been
proposed in the game theory literature to deal with coalitions
\cite{BernheimPelegWhinston,MW96}.  However, these notions do not deal
with players who act in unexpected ways.  

There can be many reasons that
players act in unexpected ways.  One, of course, is that they are indeed
irrational.  However, often seemingly irrational behavior can be
explained by players having unexpected utilities.  For example, in a
peer-to-peer network like Kazaa or Gnutella, it would seem that no
rational agent should share files. Whether or not you can get a file
depends only on whether other people share files.  Moreover, 
there are disincentives for sharing (the possibility of
lawsuits, use of bandwidth, etc.). Nevertheless, people do share
files.  However, studies of the Gnutella network have shown that almost
70 percent of users share no files and nearly 50 percent of
responses are from the top 1 percent of sharing hosts~\shortcite{AH00}.
Is the behavior of the sharing hosts irrational?  It is if we assume
appropriate utilities.  But perhaps sharing hosts get a big kick out of
being the ones that provide everyone else with the music they play.  Is
that so irrational?  In other cases, seemingly irrational behavior can
be explained by faulty computers or a faulty network (this, of course,
is the concern that work on Byzantine agreement is trying to address),
or a lack of understanding of the game.  

To give just one example of a stylized game where this issue might be
relevant, consider a group of $n$ bargaining agents.
If they all stay and bargain, then all get 2.  However, if any agent
leaves the bargaining table, those who leave get 1, while those who stay
get 0.  Clearly everyone staying at the bargaining table is a
$k$-resilient Nash
equilibrium for all $k \ge 0$, and it is Pareto optimal (everyone in
fact gets the highest 
possible payoff).  But, especially if $n$ is large, this equilibrium is
rather ``fragile''; all it takes is one person to leave the bargaining
table for those who stay to get 0.  

Whatever the reason, as pointed out by Abraham et al.~\citeyear{ADGH06},
it seems important to 
design strategies that tolerate such unanticipated behaviors, so that
the payoffs of the users with ``standard'' utilities do not get
affected by the nonstandard players using different strategies.
This can be viewed as a way of adding fault tolerance to equilibrium
notions.  To capture this intuition, Abraham et al. \cite{ADGH06} define
a strategy profile to be \emph{$t$-immune} if no player who does
\emph{not} deviate is worse off if up to $t$ players do deviate.  
Note the difference between resilience and immunity.  A strategy profile
is resilient if deviators do not gain by deviating; a profile is immune
if non-deviators do not get hurt by deviators.  In the example above,
although everyone bargaining is a $k$-resilient Nash equilibrium for all
$k \ge 0$, it is not 1-immune.

Of course, we may want to combine resilience and resilience; a strategy
is \emph{$(k,t)$-robust} if it is both $k$-resilient and $t$-immune.
(All the informal definitions here are completely formalized in
\cite{ADGH06,ADH07}.)  A Nash equilibrium is just a (1,0)-robust
equilibrium.  Unfortunately, for $(k,t) \ne (1,0)$, a $(k,t)$-robust
equilibrium does not exist in general.  Nevertheless, there are a number
of games of interest where they do exist; in particular, they can exist
if players can take advantage of a \emph{mediator}, or trusted third party.
To take just one example, consider Byzantine agreement \cite{PSL}.
Recall that in Byzantine agreement
there are $n$ soldiers, up to $t$ of which may be faulty (the $t$ stands
for {\em traitor}), one of which is the general.  The general has an
initial preference to attack or retreat. We want a protocol that
guarantees that  (1) all {\em nonfaulty\/} soldiers reach the same
decision, and 
(2) if the general is nonfaulty, then the decision is the general's
preference.  It is trivial to solve Byzantine agreement with a mediator:
the general simply sends the mediator his preference, and the mediator
sends it to all the soldiers.  

The obvious question of interest is whether we can \emph{implement} the
mediator.  That is, can the players in the system, just talking among
themselves (using what economists call ``cheap talk'') simulate the
effects of the mediator.  This is a question that has been of interest
to both the computer science community and the game theory community.
In game theory, the focus has been on whether a Nash equilibrium in a
game with a mediator can be implemented using cheap
talk (cf.~\shortcite{Barany92,Bp03,F90,Gerardi04,Heller05,UV02,UV04}).
In cryptography, the focus has been on \emph{secure multiparty
computation} \shortcite{GMW87,SRA81,yao:sc}.  
Here it is assumed that each agent $i$ has some private
information $x_i$ (such private information, like the general's
preference, is typically called the player's \emph{type} in game
theory).  Fix a function $f$.  The goal is have agent $i$
learn $f(x_1, \ldots, x_n)$ without learning anything about $x_j$ for
$j \ne i$ beyond what is revealed by the value of $f(x_1, \ldots,
x_n)$.  With a trusted mediator, this is trivial: each agent $i$ just
gives the mediator its private value $x_i$; the mediator then sends each
agent $i$ the value $f(x_1, \ldots, x_n)$.  Work on multiparty
computation provides general conditions under  which this
can be done (see \cite{goldreich03} for an overview).  
Somewhat surprisingly, despite there being over 20 years of work on this
problem in both computer science and game theory, until recently, there
has been no interaction between the communities on this topic.

Abraham et al.~\citeyear{ADGH06,ADH07} essentially characterize when
mediators can be implemented.  To understand the results,
three games need to be considered:~an
\emph{underlying game} $\Gamma$, an extension $\Gamma_d$ of $\Gamma$
with a mediator, and a cheap-talk extension $\Gamma_{CT}$ of $\Gamma$.
$\Gamma$ is assumed to be a
\emph{normal-form Bayesian game}: each player has a type from some type
space with a known distribution over 
types, and must choose an action (where the choice can depend on his
type).  The utilities of the players depend on the types and actions
taken.  For example, in Byzantine agreement, the possible types of the
general are 0 and 1, his possible initial preferences (the types of the
other players are irrelevant).  The players' actions are to attack or
retreat.  The assumption that there is a distribution over the
general's preferences is standard in game theory, although not so much
in distributed computing.  Nonetheless,  in many applications of
Byzantine agreement, it seems reasonable to assume such a distribution.
Roughly speaking, a cheap talk game \emph{implements} a game with a
mediator if it induces the same distribution over actions in the
underlying game, for each
type vector of the players. With this background, I can summarize the
results of Abraham et al.

\begin{itemize}
\item If $n > 3k + 3t$, a $(k,t)$-robust strategy $\vec{\sigma}$ with a
mediator can be implemented using cheap talk (that is, there is a
$(k,t)$-robust strategy $\vec{\sigma}'$ in the cheap talk game such that
$\vec{\sigma}$ and $\vec{\sigma}'$ induce the same distribution over
actions in the underlying game).  Moreover, the implementation
requires no knowledge of other agents' utilities, and the cheap talk
protocol has bounded running time
that does not depend on the utilities.

\item If $n \le 3k + 3t$ then, in general, mediators cannot be implemented
using cheap talk without  knowledge of other agents' utilities.
Moreover, even if other agents' utilities are
known, mediators cannot, in general, be implemented without having a
$(k+t)$-punishment strategy (that is, a strategy that, if used by all
but at most $k+t$ players, guarantees that every player gets a worse
outcome than they do with the equilibrium strategy) nor with bounded
running time. 

\item If $n > 2k+3t$, then mediators can be implemented using cheap talk
if there is a punishment strategy (and utilities are known) in
finite expected running time that does not depend on the utilities.

\item If $n \le 2k+3t$ then mediators cannot, in general, be implemented,
even if there is a punishment strategy and utilities are known.

\item If $n > 2k+2t$ and there are broadcast channels then, for all
$\epsilon$, mediators can be $\epsilon$-implemented (intuitively, 
there is an implementation where players get utility within $\epsilon$
of what they could get by deviating) 
using cheap talk, with
bounded expected running time that does not depend on the utilities.

\item If $n \le 2k+2t$ then mediators cannot, in general, be
$\epsilon$-implemented, even with broadcast channels.  Moreover, even
assuming cryptography and polynomially-bounded players, the expected
running time of an implementation depends on the utility functions
of the players and $\epsilon$.

\item If $n > k+3t$ then, assuming cryptography and polynomially-bounded
players, mediators can be $\epsilon$-implemented using cheap talk,
but if $n \le 2k + 2t$, then the running time depends on the utilities
in the game and $\epsilon$. 

\item If $n \le k+3t$, then even assuming cryptography,
polynomially-bounded players, and a $(k+t)$-punishment strategy,
mediators cannot, in general, be $\epsilon$-implemented
using cheap talk.

\item If $n > k+t$ then, assuming cryptography, polynomially-bounded
players, and a public-key infrastructure (PKI), we can
$\epsilon$-implement a mediator.
\end{itemize}

All the possibility results showing that mediators can be
implemented use techniques from secure multiparty computation.  The
results showing that that if $n \le 3k+3t$, then we cannot implement 
a mediator without knowing utilities and that, even if utilities are known, a
punishment strategy is required, use the fact that Byzantine agreement
cannot be reached if $t 
< n/3$; the impossibility result for $n \le 2k + 3t$ also uses a
variant of Byzantine agreement.   These results provide an excellent
illustration of how the interaction between computer science and game
theory can lead to fruitful insights.  Related work on  implementing
mediators can be found in \cite{GK06,HT04,IML05,KN08,LMPS04,LT06}.

\section{Taking Computation Into Account}\label{sec:computation}
Nash equilibrium does not take computation into account.  To see why
this might be a problem, consider the following example, taken from
\cite{HPass08}.  

\begin{example}\label{xam1} You are given a number $n$-bit number $x$.
You can guess whether it is prime, or play safe and say
nothing. If you guess right, you get \$10; if you guess wrong, you lose
\$10; if you play safe, you get \$1.
There is only one Nash equilibrium in this 1-player game: giving the
right answer. But if $n$ is large, this is almost certainly not what people
will do.  Even though primality testing can be done in polynomial time,
the costs for doing so (buying a larger computer, for example, or
writing an appropriate program), will probably not be worth it for most
people.  The point here is that Nash equilibrium is not taking the cost
of computing whether $x$ is prime into account.
\end{example}

There have been attempts in the game theory community to define solution
concepts that take computation into account, going back to the work of
Rubinstein \citeyear{Rub85}. (See \cite{Kalai90} for an overview of the
work in this area in the 1980s, and \cite{BKK07} for more recent work.)  
Rubinstein assumed that players choose a finite automaton to play the
game rather than choosing a strategy directly; a player's utility depends both 
on the move made by the automaton and the complexity of the automaton
(identified with the number of states of the automaton).
Intuitively, automata that use more states are seen as representing more
complicated procedures.   Rafael Pass and I \citeyear{HPass08} provide a
general game-theoretic framework that takes computation into account.
(All the discussion in this section is taken from \cite{HPass08}.)
Like Rubinstein, we view all players as choosing a machine, but we use
Turing machines, rather than finite automata.  We
associate a complexity, not just with a machine, but with the machine
and its input.  This is important in Example~\ref{xam1}, where the
complexity of computing whether $x$ is prime depends, in general, on the
length of $x$.

The complexity 
could represent the running time of or space used by the machine on that
input.  The complexity can also be used to capture the complexity of the
machine itself (e.g., the number of states, as in Rubinstein's case) or to
model the cost of searching for a new strategy to replace one that the
player already has.  
(One of the reasons that players follow a recommended strategy is
that there may be too much effort involved in trying to find a new one;
I return to this point later.)

We again consider Bayesian games, where each player has a type.  In a
standard Bayesian game, an agent's utility depends on the type profile
and the action profile (that is, every player's type, and the action
chosen by each player).  In a \emph{computational Bayesian game}, each
player $i$ chooses a Turing machine.  Player $i$'s type $t_i$ is taken to be
the input to player $i$'s Turing machine $M_i$.  The output of $M_i$ on
input $t_i$ is taken to be player $i$'s action.  There is also a
complexity associated with the pair $(M_i,t_i)$.  Player $i$'s utility
again depends on the type profile and the action profile,  
and also on the complexity profile.  The reason we consider the whole
complexity profile in determining player $i$'s utility, as opposed to
just $i$'s complexity, is that, for example, $i$ might be happy as long
as his machine takes fewer steps than $j$'s.  Given these definitions,
we can define Nash equilibrium as usual.  With this definition, by
defining the complexity appropriately, it will be the case that playing
safe for sufficiently large inputs will be an equilibrium.

Computational Nash equilibrium also gives a plausible explanation of
observed behavior in finitely-repeated prisoner's dilemma.  

\begin{example}\label{xam:pd} Recall that
prisoner's dilemma, in prisoner's dilemma, there are two prisoners, who
can choose to either cooperate or defect.
As described in the table below, if they both cooperate, they both get
3; if they both defect, then both get 1; if one defects and the other
cooperates, the defector gets 5 and the cooperator gets $-5$.
(Intuitively, the cooperator stays silent, while the defector ``rats
out'' his partner.  If they both rat each other out, they both go to jail.)
\begin{table}[htb]
\begin{center}
\begin{tabular}{c |  c c}
& $C$ & $D$\\
\hline
$C$ &(3,3) &$(-5,5)$ \\
$D$  &$(5,-5)$  &(-3,-3)  \\
\end{tabular}
\end{center}
\end{table}

It is easy to see that defecting dominates cooperating: no matter what
the other player does, a player is better off defecting than
cooperating.  Thus, ``rational'' players should defect.  And, indeed,
$(D,D)$ is the only Nash equilibrium of this game.  Although $(C,C)$
gives both players a better payoff than $(D,D)$, this is not an
equilibrium.

Now consider finitely repeated prisoner's dilemma (FRPD), where
prisoner's dilemma is played for some fixed number $N$ of rounds.  
The only Nash equilibrium is to always defect; this can be seen by a
backwards induction argument.  (The last round is like the one-shot game,
so both players should defect; given that they are both defecting at the
last round, they should both defect at the second-last round; and so on.)
This seems quite unreasonable.  And, indeed, in experiments, people do
not always defect.  In fact, quite often they cooperate throughout the
game.  Are they irrational?  It is hard to call this irrational
behavior, given that the ``irrational'' players do much better than
supposedly rational players who always defect.  There have been many
attempts to explain cooperation in FRPD in the literature (see, for
example, \cite{KMRW}).  Indeed, there have even been well-known attempts
that take computation into account; it can be shown that
if players are restricted to 
using a finite automaton with bounded complexity, then there exist
equilibria that allow for cooperation \cite{Ney85,PY94}.
However, the strategies used in those equilibria are quite complex, and
require the use of large automata;
as a consequence this approach does not seem to provide a satisfactory
explanation as to why people choose to cooperate.

Using the framework described above leads to a straightforward explanation.  
Consider the \emph{tit-for-tat} strategy, which proceeds as follows: a
player cooperates at the first round, and then at round $m+1$, does
whatever his opponent did at round $m$.  Thus, if the opponent cooperated
at the previous round, then you reward him by continuing to cooperate; if he
defected at the previous round, you punish him by defecting.  If both players
play tit-for-tat, then they cooperate throughout the game.  Interestingly,
tit-for-tat does exceedingly well in FRPD tournaments, where computer
programs play each other \cite{Axelrod}.  

Tit-for-tat is a simple program, which needs very little memory.  
Suppose that 
we charge even a modest amount for memory usage, and that there is a
discount factor $\delta$, with $.5 < \delta < 1$, so that if the player
gets a reward of $r_m$ in round $m$, his total reward over the whole
$N$-round game is taken to be $\sum_{m=1}^N  \delta^m r_m$.  In this
case, it is easy to see that, 
no matter what the 
cost of memory is, as long as it is positive, for a sufficiently long
game, it will be a Nash equilibrium for both players to play
tit-for-tat.  For the best response to tit-for-tat is to play
tit-for-tat up to the last round, and then to defect.  But following
this strategy requires the player to keep track of the round number,
which requires the use of extra memory.  The extra gain of \$2 achieved
by defecting at the last round, if
sufficiently discounted, will not be worth the cost of keeping track of
the round number.  

Note that even if only one player is computationally bounded and is
charged for memory, and memory is free for the other player, then there
is a Nash equilibrium where the bounded player plays tit-for-tat, while
the other player plays the best response of cooperating up (but not
including) to the round of
the game, and then defecting.
\end{example}

Although with standard games there is always a Nash equilibrium, this
is not the case when we take computation into account, as the following
example shows.

\begin{example}\label{xam:roshambo}
Consider roshambo (rock-paper-scissors).
We model playing rock, paper, and scissors as playing 0, 1, and 2,
respectively.  The payoff to player 1 of the outcome $(i, j)$ is 1 if $i
= j \oplus 1$ (where $\oplus$ denotes addition mod 3), $-1$ if $j = i
\oplus 1$, and 0 if $i = j$.  Player 2's playoffs are the negative of those
of player 1; the game is a zero-sum game.  As is well known, the unique
Nash equilibrium of this game has the players randomizing uniformly
between 0, 1, and 2. 

Now consider a computational version of roshambo.  Suppose that we 
take the complexity of a deterministic strategy to be 1, and the
complexity of 
a strategy that uses randomization to be 2, and take player $i$'s
utility to be his payoff in the underlying Bayesian game minus the complexity
of his strategy. 
Intuitively, programs involving randomization are more
complicated than those that do not randomize.  With this utility
function, it is easy to see that there is no Nash equilibrium.  For
suppose that $(M_1, M_2)$ is an equilibrium.  If $M_1$ uses randomization,
then 1 can do better by playing the deterministic strategy $j\oplus 1$,
where $j$ is the action that gets the highest probability according to
$M_2$ (or is the deterministic choice of player 2 if $M_2$ does not use
randomization).  Similarly, $M_2$ cannot use randomization.  But it is well
known (and easy to check) that there is no equilibrium for roshambo with 
deterministic strategies. 
\end{example}

Is the lack of Nash equilibrium a problem?  Perhaps not.  Taking
computation into account should cause us to rethink things.  In
particular, we may want to consider other solution concepts.  But, as
the examples above show, Nash equilibrium does seem to make reasonable
predictions in a number of games of interest.  Perhaps of even more
interest, using computational Nash equilibrium lets us provide a
game-theoretic account of security.

The standard framework for multiparty security 
does not take into account whether players have an incentive 
to execute the protocol. That is, if there were a trusted
mediator, would player $i$ actually use the recommended protocol
even if $i$ would be happy to use the services of the mediator to
compute the function $f$?  Nor does it 
take into account whether the adversary has an incentive to undermine
the protocol.   

Roughly speaking, the game-theoretic definition says that $\Pi$ is a
\emph{game-theoretically secure} (cheap-talk) protocol for computing $f$
if, for all choices of the utility function,
if it is a Nash equilibrium to play with the mediator to
compute $f$, then it is also a Nash equilibrium to use $\Pi$ to compute
$f$.  Note that this definition does not mention privacy.  It does not
need to; this is taken care of by choosing the utilities appropriately.
Pass and I \citeyear{HPass08} show that, under minimal assumptions, this
definition is essentially equivalent to  a variant of \emph{zero knowledge}
\cite{GMR} called \emph{precise zero knowledge} \cite{MP06}.
Thus, the two approaches used for dealing with ``deviating'' players
in two game theory and cryptography---\emph{Nash equilibrium} 
and \emph{zero-knowledge ``simulation''}---are intimately
connected; indeed, they are essentially equivalent once we take
computation into account appropriately.  

\section{Taking (Lack of) Awareness Into Account}\label{sec:awareness}

Standard game theory models implicitly assume that all significant
aspects of the game (payoffs, moves available, etc.) are common
knowledge among the players. 
However, this is not always a reasonable assumption. For example,
sleazy companies assume that consumers are not aware that they can
lodge complaints if there are problems; in a war setting, having
technology that an enemy is unaware of (and thus being able to make
moves that the enemy is unaware of) can be critical; in financial
markets, some investors may not be aware of certain investment
strategies (complicated hedging strategies, for example, or
tax-avoidance strategies).

\begin{figure}[ht]
\centering \epsfxsize=9cm \epsffile{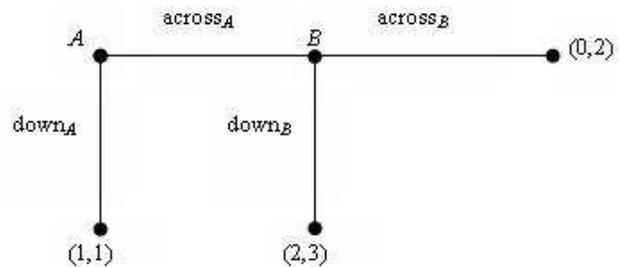}
\caption{A simple game.}\label{fig:game1}
\end{figure}

To understand the impact of adding the possibility of unawareness to
the analysis of games, consider the game shown in
Figure~\ref{fig:game1} (this example, and all the discussion in this
section, is taken from \cite{HR06}).  One Nash equilibrium of this game
has $A$ playing across$_A$ and $B$ 
playing down$_B$.  However, suppose that $A$ is not aware that $B$
can play down$_B$.  In that case, if $A$ is rational, $A$ will play
down$_A$. Therefore, Nash equilibrium does not seem to be the
appropriate solution concept here. Although $A$ would play
across$_A$ if $A$ knew that $B$ were going to play down$_B$, $A$
cannot even contemplate this possibility, let alone know it.

To find an appropriate analogue of Nash equilibrium in games where
players may be unaware of some possible moves, we must first find an
appropriate representation for such games. The first step in doing
so is to explicitly represent what players are aware of at each
node.  We do this by using what we call an \emph{augmented game}.

Recall that an \emph{extensive game} is described by a \emph{game tree}. 
Each node in the tree describes a partial
\emph{history} of the game---the sequence of moves that led to that
node. Associated with each node is the player that moves at that
node. Some nodes where a player $i$ moves are grouped together into
an \emph{information set for player $i$}.  
Intuitively, if player $i$ is at some node in an information set $I$,
then $i$ does not know which  node of $I$ describes the true situation;
thus, at all nodes in $I$, $i$ must make the same move.
An augmented game is an extensive game with one more feature:
associated with each node in the game tree where player $i$ moves is the
\emph{level of 
awareness} of player $i$---the set of histories that player $i$ is aware
of.  

We use the player's awareness level as a way of keeping track of how the
player's awareness changes over time.
For example, perhaps $A$ playing across$_A$ will result in $B$
becoming aware of the possibility of playing down$_B$. In financial
settings, one effect of players using certain investment strategies
is that other players become aware of the possibility of using that
strategy. Strategic thinking in such games must take this
possibility into account.  We would model this possibility by having
some probability of $B$'s awareness level changing.  (The formal
definition of augmented game can be found in \cite{HR06}.)

For example, suppose that in the game shown in Figure~\ref{fig:game1}
\begin{itemize}
\item players $A$ and $B$ are aware of all histories of the game;
\item player $A$ is uncertain as to  whether player $B$ is aware of run
$\<$across$_A$,\,down$_B\>$ and believes that he is unaware of it
with probability $p$; and
\item the type of player $B$ that is
aware of the run $\<$across$_A$, down$_B\>$ is aware that player $A$
is aware of all histories, and he knows $A$ is uncertain about his
awareness level and knows the probability $p$.
\end{itemize}

\begin{figure}[htb]
\centering \epsfxsize=9cm \epsffile{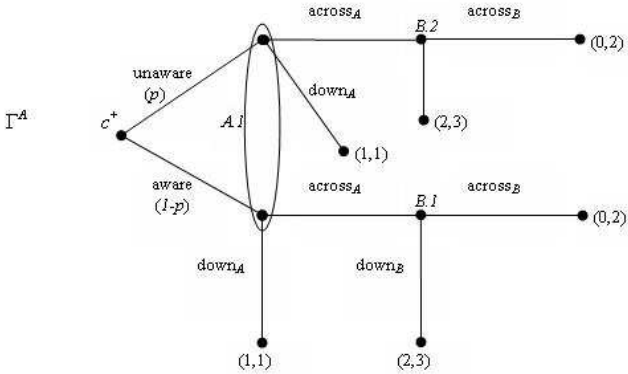} \caption{The
augmented game $\Gamma^A$.} \label{fig:A1game}
\end{figure}

Because $A$ and $B$ are actually aware of all histories of the underlying
game, from the point of view of the modeler, the augmented game is
essentially identical to the game described in
Figure~\ref{fig:game1}, with the awareness level of both players $A$
and $B$ consisting of all histories of the underlying game. However, when
$A$ moves at the node labeled $A$ in the modeler's game, she
believes that the actual augmented game  is $\Gamma^A$, as described
in Figure~\ref{fig:A1game}.  In $\Gamma^A$, nature's initial move
captures $A$'s uncertainty about $B$'s awareness level. At the
information set labeled $A.1$, $A$ is aware of all the runs of the
underlying game. Moreover, at this information set, $A$ believes
that the true game is $\Gamma^A$.

\begin{figure}[htb]
\centering \epsfxsize=7cm \epsffile{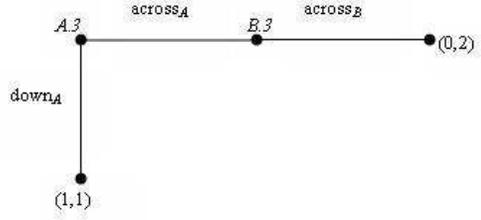} \caption{The
augmented game $\Gamma^B$.} \label{fig:A2game}
\end{figure}

At the node labeled $B.1$, $B$ is aware of all the runs of the
underlying game and believes that the true game is the modeler's
game; but at the node labeled $B.2$, $B$ is not aware that he can
play down$_B$, and so believes that the true game is the augmented
game $\Gamma^B$ described in Figure~\ref{fig:A2game}. At the nodes
labeled $A.3$ and $B.3$ in the game $\Gamma^B$, neither $A$ nor $B$
is aware of the move down$_B$.  Moreover, both players think the
true game is $\Gamma^B$.

As this example should make clear, to model a game with possibly
unaware players, we need to consider, not just one augmented game,
but a collection of them.  Moreover, we need to describe, at each
history in an augmented game, which augmented game the player
playing at that history believes is the actual augmented game being
played.

To capture these intuitions, starting with an underlying extensive-form
game $\Gamma$, we define a \emph{game with awareness
based on $\Gamma$} 
to be a tuple $\Gamma^* = (\G, \Gamma^m, \F)$, where
\begin{itemize}
\item $\G$ is a countable set of augmented games based on $\Gamma$, of
which one is
$\Gamma^m$;
\item $\F$ maps an augmented game $\Gamma^+ \in\G$ and a history $h$ in
$\Gamma^+$ such that $P^+(h)=i$ to a pair $(\Gamma^h,I)$, where
$\Gamma^h\in \G$ and $I$ is an information set for player $i$ in game
$\Gamma^h$.
\end{itemize}
Intuitively, $\Gamma^m$ is the game from the point of view of an
omniscient modeler. If player $i$ moves at $h$ in game $\Gamma^+ \in
\G$ and $\F(\Gamma^+,h) = (\Gamma^h,I)$, then $\Gamma^h$ is the game
that $i$ believes to be the true game when the history is $h$, and $I$
consists of the set of histories  in $\Gamma^h$ he currently
considers possible. For example, in the examples described in
Figures~\ref{fig:A1game} and~\ref{fig:A2game}, taking $\Gamma^m$ to
the augmented game in Figure~\ref{fig:game1}, we have
$\F(\Gamma^m,\<\, \>) = (\Gamma^A,I)$, where $I$ is the information
set labeled $A.1$ in Figure~\ref{fig:A1game}, and
$\F(\Gamma^A,\<$unaware,across$_A\>) =
(\Gamma^B,\{\<$across$_A\>\})$.
There are a number of consistency conditions that have to be satisfied
by the function $\F$; the details can be found in \cite{HR06}.

The standard notion of Nash equilibrium consists of a profile of
strategies, one for each player.  Our generalization consists of a
profile of strategies, one for each pair $(i,\Gamma')$, where
$\Gamma'$ is a game that agent $i$ considers to be the true game in
some situation. Intuitively, the strategy for a player $i$ at
$\Gamma'$ is the strategy $i$ would play in situations where $i$
believes that the true game is $\Gamma'$. To understand why we may
need to consider different strategies consider, for example, the
game of Figure~\ref{fig:game1}. $B$ would play differently depending
on whether or not he was aware of down$_B$. Roughly speaking, a profile
$\vec{\sigma}$ 
of strategies, one for each pair $(i,\Gamma')$, is a
\emph{generalized Nash equilibrium} if $\sigma_{i,\Gamma'}$
is a best response for player $i$ if the true game is
$\Gamma'$, given the strategies $\sigma_{j,\Gamma'}$ being used by the
other players in 
$\Gamma'$.  As shown in \cite{HR06}, every game with awareness has a
generalized Nash equilibrium.

A standard extensive-form game $\Gamma$ can be viewed as a special case
of a game with awareness, 
by taking $\Gamma^m = \Gamma$, $\G = \{\Gamma^m\}$, and $\F(\Gamma^m,h)
= (\Gamma^m,I)$, where $I$ is the information set that contains $h$.
Intuitively, $\Gamma$ corresponds to the game of awareness where it is
common knowledge that $\Gamma$ is being played.  We call this the
\emph{canonical representation} of $\Gamma$ as a game with awareness.
It is not hard to show
that a strategy profile $\vec{\sigma}$ is a Nash equilibrium of $\Gamma$ iff
it is a generalized Nash equilibrium of the canonical representation of
$\Gamma$ as a game with awareness.  Thus, generalized Nash equilibrium
can be viewed as a generalization of standard Nash equilibrium.

Up to now, I have considered only games 
where players are not aware of their lack of awareness.  But in some
games, a player might be aware that there are
moves that another player (or even she herself) might be able to
make, although she is not aware of what they are. Such awareness of
unawareness can be quite relevant in practice. For example, in a
war setting, even if one side cannot conceive of a
new technology available to the enemy, they might believe that there
is some move available to the enemy without understanding what that
particular move is.  This, in turn, may encourage peace overtures.
To take another example, an agent might delay making a decision
because she considers it possible that she might learn about more
possible moves, even if she is not aware of what these moves are.

Although, economists usually interpret awareness as ``being able to
conceive about an event or a proposition'', there are other possible
meanings for this concept. For example, awareness may
also be interpreted as ``understanding the primitive concepts in an
event or proposition'', or as ``being able to determine if an event
occurred or not'', or as ``being able to compute the consequences of
some fact'' \cite{FH}.
If we interpret ``lack of awareness'' as ``unable to compute'' (note
that this interpretation is closely related to the discussion of the
previous section!), then awareness of unawareness becomes even more
significant.  Consider a chess game.  Although all players
understand in principle all the moves that can be made, they are
certainly not aware of all consequences of all moves.   A more
accurate representation of chess would model this computational
unawareness explicitly. We provide such a representation.

Roughly speaking, we capture the fact that player $i$ is aware that,
at a node $h$ in the game tree, there is a move that $j$ can make
she ($i$) is not aware by having $i$'s subjective representation of
the game include a ``virtual'' move for $j$ at node $h$.  Since $i$
might have only an incomplete understanding of what can happen after
this move,
$i$ simply describes what she believes will be the game after the
virtual move,
to the extent that she can.  In particular, if she has no idea what will
happen after the virtual move, then she can describe her beliefs
regarding the payoffs of the game.
Thus, our representation can be viewed as a generalization of how chess
programs
analyze chess games.  They explore the game tree up to a certain
point, and then evaluate the board position at that point. We can
think of the payoffs following a virtual move by $j$ in $i$'s
subjective representation of a chess game as describing the
evaluation of the board from $i$'s point of view. This seems like a
much more reasonable representation of the game than the standard
complete game tree!  

All the definitions of games with awareness can be generalized to
accommodate awareness of unawareness.  In particular, we can
define a generalized Nash equilibrium as
before, and once again show that every game with awareness (including
awareness of unawareness) has a generalized Nash equilibrium
\cite{HR06}.

There has been a great deal of work recently on modeling unawareness in
games.  The first papers on the topic was by Feinberg
\citeyear{Feinberg04,Feinberg05}.  My work with R\^{e}go \citeyear{HR06}
was the first to consider awareness in extensive games, modeling how
awareness changed over time.   There has been a recent flurry on the
topic in the 
economics literature; see, for example, \cite{HMS06,LI06,LI06a,Ozbay06}.
Closely related is work on logics that include awareness.  This work
started in the computer science literature \cite{FH}, but more
recently, the bulk of the work has appeared in the economics literature
(see, for example, \cite{DLR98,Hal34,HR05,HMS03,MR94,MR99}).

\section{Conclusions}
I have considered three ways of going beyond standard Nash equilibrium,
which take fault tolerance, computation, and lack of awareness into
account, respectively.  These are clearly only first steps.  Here are
some directions for further research (some of which I am currently
engaged in with my collaborators):
\begin{itemize}
\item For example, while $(k,t)$-robust equilibrium does seem to be a
reasonable way of capturing some aspects of robustness, for some
applications, it does not go far enough.  I said earlier that in
economics, all players were assumed to be strategic, or ``rational''; in
distributed computing, all players were either ``good'' (and followed
the recommended protocol) or ``bad'' (in which case they could be
arbitrarily malicious).  Immunity takes into account the bad players.
The definition of immunity requires that the rational players are not
hurt no matter what the ``bad'' players do.  But this may be too
strong.  As Ayer et al.~\citeyear{Alvisi05} point out, it is reasonable
to expect a certain fraction of players in a system to be ``good'' and
follow the recommended protocol, even if it is not a best reply.  
In general, it may be hard to figure out what the best reply is, so if
following the recommended protocol is not unreasonable, they will do
that.  (Note that this can be captured in a computational model of
equilibrium, by charging for switching from the recommended strategy.)

There may be other standard ways that
players act irrational.  For example, Kash, Friedman, and
I~\citeyear{KFH07} consider scrip systems, where players perform
work in exchange for scrip.  There is a Nash equilibrium where everyone
uses a \emph{threshold strategy},  performing work only when they have
less scrip than some threshold amount.  Two standard ways of acting
``irrationally'' in such a system are to (a) hoard scrip and (b) provide
service for free (this is the analogue of posting music on
Kazaa).  A robust solution should take into account these more standard
types of irrational behavior, without perhaps worrying as much about
arbitrary irrational behavior.  
\item The definitions of computational Nash equilibrium considered only
Bayesian games.  What would appropriate solution concepts be for
extensive-form games?  Some ideas from the work on awareness seem
relevant here, especially if we think of ``lack of awareness'' as
``unable to compute''.
\item Where do the beliefs come from in an equilibrium with awareness?
That is, if I suddenly become aware that you can make a certain move,
what probability should I assign to you making that move?  Ozbay
\citeyear{Ozbay06} proposes a solution concept where the beliefs are
part of the solution concept.  He considers only a simple setting, where
one player is aware of everything (so that revealing information is
purely strategic).  Can his ideas be extended to a more general setting?
\end{itemize}

Agents playing a game can be viewed participating in a concurrent,
distributed protocol.  Game theory does not take the asynchrony into
account, but it can make a big difference.  For example, all the results
from \cite{ADGH06,ADH07} mentioned in Section~\ref{sec:robust} depend on
the system being synchronous.  Things are more complicted in
asynchronous settings.  Getting solution concepts and that deal
well with with asynchrony is clearly important.  

Another issue that plays a major role in computer science but has thus
far not been viewed as significant in game theory, but will, I believe,
turn out to be important to the problem of defining appropriate solution
concepts, is the analogue of specifying and verifying programs.    Games
are typically designed to solve 
certain problems.  Thus, for example, economists want to design a
spectrum auction so that the equilibrium has certain features.  
As I pointed out in an earlier overview \cite{Hal38}, 
game theory has typically focused on ``small'' games: games that are
easy to describe, such as Prisoner's Dilemma.  The focus has
been on subtleties regarding  basic issues such as rationality and
coordination.  To the extent that game theory is used to tackle
larger, more practical problems, and especially to the extent that
it is computers, or software agents, playing games, rather than people,
it will be important to specify
carefully exactly what a solution to the game must accomplish.
For example,  in the context of a
spectrum auction, a specification will have to address what should
happen if a computer crashes while an agent is 
in the middle of transmitting a bid, how to deal with agents bidding on
slow lines, dealing with agents who win but then go bankrupt, and so on.

Finding logics to reason about solutions, especially doing so in a way  
that takes into account robustness and asynchrony, seems to me a
difficult and worthwhile challenge.  Indeed, one desideratum for a good
solution concept is that it should be easy to reason about. 
Pursuing this theme, computer scientists have learned that one good way
of designing correct 
programs is to do so in a modular way.  Can a similar idea be applied in
game theory?  That is, can games designed for solving smaller problems be
combined in a seamless way to solve a larger problem.  If so, results
about \emph{composability of solutions} will be 
needed; we might want a solution concept that allows for such
composability.

\bibliographystyle{chicagor}

\end{document}